\documentclass[usenatbib]{mn2e}

\usepackage{epsfig}
\usepackage{graphicx}
\usepackage{tabularx}
\usepackage{amsmath}
\usepackage{amssymb}
\usepackage{amstext}
\usepackage{amsfonts}

\newcommand{\bm}[1]{\ensuremath\mathbf{#1}}
\newcommand{\avrg}[1]{\ensuremath{\langle #1 \rangle}}
\newcommand{\dd}{\ensuremath\text{d}}
\newcommand{\pder}[2]{\ensuremath\frac{\partial #1}{\partial #2}}
\newcommand{\der}[2]{\ensuremath\frac{\dd #1}{\dd #2}}
\title{Stochastic heating of cooling flows}

\author[Georgi Pavlovski, Edward Pope]%
{Georgi Pavlovski$^1$, Edward C.D. Pope$^2$\\
$^1$School of Physics and Astronomy, University of Southampton,
Southampton SO17 1BJ, U.K. gbp@phys.soton.ac.uk\\
$^2$School of Physics and Astronomy, University of Victoria, BC, V8P 1A1,
Canada}
\date{Accepted .... . Received .... ; in original form .... }

\begin{document}

\maketitle

\begin{abstract}
It is generally accepted that the heating of gas in clusters of
galaxies by active galactic nuclei (AGN) is a form of
feedback. Feedback is required to ensure a long term, sustainable
balance between heating and cooling.  This work investigates the
impact of proportional stochastic feedback on the energy balance in
the intracluster medium. Using a generalised analytical model for a
cluster atmosphere, it is shown that an energy equilibrium can be
reached exponentially quickly. Applying the tools of stochastic
calculus it is demonstrated that the result is robust with regard to
the model parameters, even though they affect the amount of
variability in the system.
\end{abstract}


\section{Introduction}
\label{intro}
The cooling time of gas in the cores of many galaxy clusters is much
shorter than the Hubble time.  In the absence of heat sources this gas
will cool and flow towards the centre of the cluster. However, high
resolution X-ray spectroscopy has shown that the rate at which gas
cools to low temperatures is much lower than initially expected
\citep{Peterson01,Tamura01,Peterson03}, suggesting that the gas is
being reheated.

Feedback from the active galactic nucleus (AGN) in the central galaxy
of the cluster has been considered the most promising mechanism for
the reheating of the cooling flow \citep[see, e.g., review by][and
  references therein]{McNamara07}.  Modelling of the heating process
is difficult because the physical properties of the AGN-cluster
interaction and AGN accretion are far from clear.  It has been
suggested that the cluster gas is heated by outflows, bubbles, sound
waves, thermal conduction, turbulence, and (or) a combination of the
above mentioned processes \citep[see,
  e.g.,][]{Heinz06,Brueggen02,Ruszkowski04,Voigt02,Fujita04b,Scannapieco08}. Even
though it is not clear how effective these individual processes are in
raising temperature of the centre of the cluster, it is clear that the
energy deposited by the AGN must be thermalised so that on average the
heating and the cooling rates remain equal (or at least comparable).

Currently, the theoretical description of AGN activity in cooling flow
clusters is based primarily on results from numerical models.  A
typical numerical experiment would assume a
\mbox{(magneto-)hydrodynamical} model for the intracluster plasma, and
find a detailed numerical solution of the appropriate equations.  In
this framework the relative importance of various specific physical
processes can be evaluated and their role in the heating of cooling
flows can be consequently assessed. It is prohibitively expensive,
however, to cover a large parameter space in the simulations, and
accumulation of statistics is often very difficult.  While the
simulations shed light on some relevant physical phenomena, they are
not well suited for answering questions regarding long-term stability,
variability, and statistics of the population of clusters.

In this article, an analytical model for the atmosphere of a cluster
is constructed to study the stability and evolution of the energy
balance driven by AGN feedback.  The model behaviour is expressed in
terms of stochastic differential equations, which are widely used in
studies of fluctuating phenomena, but can appear somewhat obscure at
first glance. Consequently, the model is deliberately chosen to be
very simple, so that the interpretation of the solutions of the model
equations remains robust.  This analytical approach is complementary
to the more common detailed numerical simulations. It helps to probe
otherwise difficult to reach parameter space, and addresses some of
the above mentioned questions regarding AGN feedback in clusters. Our
approach to these questions in the current work is closely related to
the ideas developed by \cite{Pope07}.  The proposed answers are
statistical in nature, which can be useful in selecting parameters for
simulations as well as interpreting their results.

Methods of stochastic calculus have been used extensively in
application to problems of statistical physics, chemistry,
computational biology and finance. In astrophysics they are mainly
used in astroparticle research \citep[see, e.g.,][]{Litvinenko09} and
are quite suitable for investigating parameter space in the context of
phenomenological models, as demonstrated in the present work.

The outline of the article is as follows.  Section~\ref{sec:toy}
presents the main model assumptions and derives the equations for
deterministic energy balance in the ICM.  Section~\ref{sec:ito}
introduces the concept of the stochastic mass deposition rate,
discusses briefly the rules of the stochastic calculus, and derives
stochastic differential equation for the heating of the
ICM. Section~\ref{sec:heat}, using a further simplification, solves
the equation, derives formulae for the mean, the probability density
function of the heating process and its asymptotic form. Section
\ref{sec:discuss} discusses possible observational implications and
limitations of the presented model.

\section{Toy model}
\label{sec:toy}

Consider a relaxed cluster of galaxies.  For simplicity the
intracluster medium (ICM) in this model is assumed to be an
unmagnetised, high temperature gas, confined by the gravitational
potential of the dark matter halo, which preserves its spherically
symmetric distribution.

\begin{figure}
\centering\includegraphics[width=0.95\linewidth]{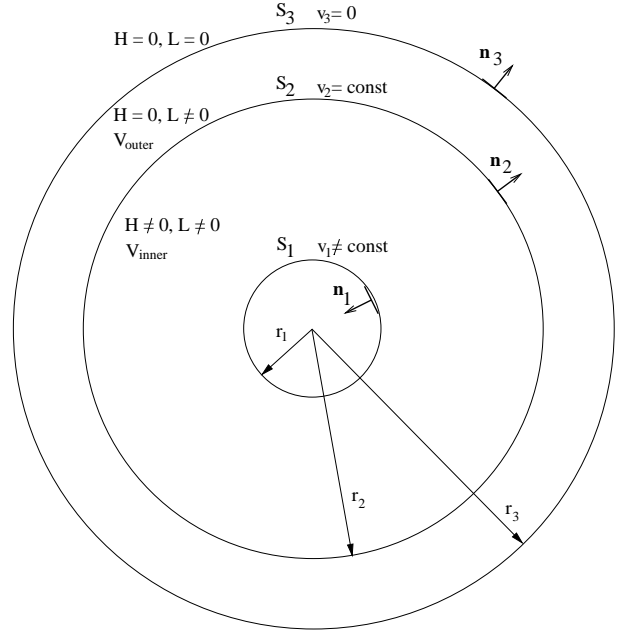}
\caption{Division of the cluster into the inner and the outer regions.  The
following notations are used: $H$ is the integral heating rate, $L$ is the
integral cooling rate, $r_i$ ($i=1,2,3$) are the radii of the spheres, $S_i$ are
the surfaces of the spheres, $\bm{n}_i$ are the normal vectors, $V_\text{inner}$
is the volume between the surfaces $S_1$ and $S_2$, $V_\text{outer}$ is the
volume between the surfaces $S_2$ and $S_3$.  The orientation of the normal
vectors $\bm{n}_1$ and $\bm{n}_3$ is external to the considered volumes (note
that $\bm{n}_1$ is external to $V_\text{inner}$, which does not extend all the
way to the centre), whereas $\bm{n}_2$ is external for $V_\text{inner}$ and
internal for $V_\text{outer}$.}
\label{fig:shells}
\end{figure}

The surface $S_3$ of the sphere with radius $r=r_3$ (see
Fig.~\ref{fig:shells}) is considered to be the outer boundary of the
ICM, beyond which the gas is no longer gravitationally bound to the
cluster. The radial velocities and the densities in the region $r>r_3$
are assumed to be negligible, $v_3=v(r\geq r_3)\approx0$,
$\rho=\rho(r\geq r_3)\approx0$.  The density of the ICM rises towards
the centre of the cluster, and the high temperature ICM cools by
emitting X-rays.  The cooling of that gas results in the classical
cooling flow -- a spherically symmetric flow towards the centre of the
cluster. The surface $S_2$ of the sphere with radius $r=r_2$ is
located sufficiently far away from the centre of the cluster so that
the density and the temperature of the gas in the volume
$V_\text{outer}$ remains unaffected by the AGN activity.  Therefore,
the absolute value of the velocity of the cooling flow on this surface
$v_2=v(r_2)$ is determined only by the cooling rate in the volume
$V_\text{outer}$, and remains approximately constant.  The volume
$V_\text{inner}$ is heated with the integral rate $H(t)\neq0$, while
it also cools with the integral rate $L_\text{inner}(t)\neq0$. In
general, values of the velocity $v_1=v(r_1,t)$, the density
$\rho_1=\rho(r_1,t)$, and the pressure $p_1=p(r_1,t)$ at $S_1$ are not
constant because of the AGN activity.  The region inside $S_1$ is
excluded from the model.

It is straightforward to write the energy balance equations for the volumes 
$V_\text{outer}$ and $V_\text{inner}$ using the generic equation,
\begin{equation}
\label{eq:toy1}
\der{E_V}{t}=-\oint_{\partial V} \left(\rho\frac{v^2}{2} + \rho\omega + 
\rho\psi\right)\bm{v}.\dd \bm{S},
\end{equation}
where $E_V$ is the sum of kinetic, thermal and potential energies of the gas
inside the volume $V$, $\omega$ is the enthalpy of the gas, $\psi$ is the
gravitational potential, $\dd\bm{S}$ is the element of the surface with an
external normal.  If the velocity of the gas is subsonic almost everywhere, then
$v^2/2<\omega$, and we can neglect the contribution of the kinetic energy term in
the equation (\ref{eq:toy1}).  Energy balance for the volume $V_\text{outer}$ is
then given by,
\begin{equation}
\label{eq:toy2}
-L_\text{outer}=-(\omega_2 +\psi_2)\dot{M}_2
\end{equation}
where $\dot{M}_2=4\pi r_2^2\rho_2v_2$ is the constant mass deposition
rate through $S_2$.  For the volume $V_\text{inner}$, the energy
balance equation can be written as
\begin{equation}
\label{eq:toy3}
\begin{split}
H(t)-L_\text{inner}(t)
&=-(\omega_1(t) +\psi_1)\dot{M}_1(t)+(\omega_2 +\psi_2)\dot{M}_2\\
&=-(\omega_1(t) +\psi_1)\dot{M}_1(t)+L_\text{outer},
\end{split}
\end{equation}
where $\dot{M}_1=4\pi r_1^2\rho_1(t)v_1(t)$ is the time dependent mass deposition
rate through $S_1$.  Differentiation of both sides of the equation
(\ref{eq:toy3}) gives,
\begin{equation}
\label{eq:toy4}
\begin{split}
\der{H}{t} + \der{L_\text{inner}}{t}
=&-(\omega_1(t) + \psi_1)\der{\dot{M}_1}{t} - \der{\omega_1}{t}\dot{M}_1(t)\\
=&(H(t) - L_\text{inner}(t) - L_\text{outer})\frac{1}{\dot{M}_1}\der{\dot{M}_1}{t}\\ 
&- \der{\omega_1}{t}\dot{M}_1(t),
\end{split}
\end{equation}
where, in the last equation, the term $-(\omega_1+\psi_1)$ was
expressed from the equation (\ref{eq:toy3}).  It is reasonable to
presume that $\dot{H}\gg\dot{L}_\text{inner}$, as the main contribution
to the heating rate integral $H$ is likely to come from a small volume
of gas close to $S_1$, in the vicinity of the AGN.  Whereas the
integral cooling rate function has the entire volume of
$V_\text{inner}$ as a support, and, therefore, is likely to be less
affected by the changes close to $S_1$.  Making this assumption
results in the following differential equation for the heating of the
ICM,
\begin{equation}
\label{eq:toy5}
\der{H}{t} =
[H(t) - L(t)]\frac{1}{\dot{M}}\der{\dot{M}}{t} - \der{\omega}{t}\dot{M}(t),
\end{equation}
where $L(t)=L_\text{outer}+L_\text{inner}(t)$, and subscripts `1' were dropped
to simplify the notation.

If the heating matches the cooling in the inner region
$H=L_\text{inner}=\text{const}$, then equation (\ref{eq:toy5}) reduces to,
 \begin{equation}
\label{eq:toy6}
-L_\text{outer}\frac{1}{\dot{M}^2}\der{\dot{M}}{t} = \der{\omega}{t},
\end{equation}
which is easily integrable,
\begin{equation}
\label{eq:toy7}
\omega(t) = \omega(t_0) 
+ L_\text{outer}\left[\frac{1}{\dot{M}(t)}-\frac{1}{\dot{M}(t_0)}\right].
\end{equation}
In this case the enthalpy is inversely proportional to the rate of
mass inflow.  It is also straightforward to get this result directly
from the equation (\ref{eq:toy3}).

\section{Stochastic feedback}
\label{sec:ito}

AGN feedback process is thought to recycle the rest-mass energy of the
infalling material during the course of its accretion onto the central
supermassive black hole. Details of the physics of accretion, methods
of the energy transfer and thermalisation in the ICM are far from
clear.  Nevertheless, as the end product of a feedback cycle, the
enthalpy of the ICM must rise and $\dot{M}$ must be reduced.  The
magnitude of the change of the mass flux, $\dd\dot{M}$, is
proportional to the magnitude of the AGN ``response'', which in turn
is likely to be proportional to the ``input signal'' $\dot{M}$,
\begin{equation}
\label{eq:ito8}
\dd\dot{M}\propto\dot{M}.
\end{equation}

Such linear relation between the response and the input signal is
found, e.g., in transient X-ray binaries, where the power of the
outburst was found to be proportional to the amount of the material in
the disk \citep{Shahbaz98}.  Although the physical processes
responsible for this relationship in X-ray binaries are likely to be
quite different from the physics of AGN in clusters, it does suggest
that the linear relation between response and the input may be a
reasonable assumption.

The proportionality coefficient in the equation (\ref{eq:ito8}) is
unlikely to be a universal constant. In principle, magnitudes of the
responses can vary given the same input $\dot{M}$.  Across a
population of clusters such variation could be due to the differences
in masses and spins of the central black holes. In an individual
cluster, the constant of proportionality could also vary because of
changes in the state of the accretion disk.

In order to account for a range of possible responses, the value of $\dd\dot{M}$
(and, therefore, $\dot{M}$) can be considered to be a random variable
parameterised by time, i.e., a stochastic process.  The equations of the toy
model in combination with a suitable distribution for $\dot{M}$ can then be used
to infer properties of the AGN heating.  The most unbiased choice of the
distribution, which still reflects the property of the proportionality
(\ref{eq:ito8}) is a uniform distribution of $\dd\dot{M}/\dot{M}$. The uniform
distribution corresponds to the ``white-noise'' stochastic process $\zeta(t)\dd
t$: $\langle\zeta(t)\rangle=0$, $\langle\zeta(t)\zeta(t')\rangle=\delta(t-t')$.
Using notations accepted in the theory of stochastic differential equations
\citep[see, e.g.,][for an introduction into the theory and applications of
stochastic calculus]{Gardiner04} this choice of the distribution implies the
following stochastic differential equation (SDE),
\begin{equation} 
\label{eq:ito10}
\frac{\dd\dot{M}}{\dot{M}}=b\dd W(t),
\end{equation}
where $b$ is a constant, and $W(t)$ is a Wiener process.

The Wiener process is defined as a continuous time random walk
(Brownian motion) in the limit of infinitesimally small step
size. According to the rules of stochastic (also called It\^o)
calculus the equality $\zeta(t)\dd t=\dd W(t)$ is interpreted
symbolically as defining the integral relation
$W(t)-W(t_0)=\int^t_{t_0}\zeta(t')\dd t'$, because strictly speaking
the derivative of $W(t)$ does not exist.  The simplest way to find an
analytical solution to a SDE is to make a suitable variable
substitution.  If the substitution transforms the term $f(\cdot)\dd
W(t)$, where $f(\cdot)$ is a function, into $c\dd W(t)$, where $c$ is
a constant, the equation can be easily integrated. According to the
rules of stochastic calculus the change can be formally done by
expanding the new variable to the second order, and using the
following identities: $(\dd W(t))^2=\dd t$, $(\dd t)^2=0$, $\dd
W(t)\dd t=0$, the first of which simply states that the variance of
Brownian motion is equal to the elapsed time.

Expansion to second order of the function $\mu=\ln\dot{M}$ relating to
the stochastic process $\dot{M}$ is given by,
\begin{equation}
\label{eq:ito11}
\begin{split}
\dd\mu &= \frac{1}{\dot{M}}\dd\dot{M}-\frac{1}{2\dot{M}^2}(\dd\dot{M})^2\\
&= b\dd W(t) - \frac{1}{2}b^2\dd t.
\end{split}
\end{equation}
The last equation can be straightforwardly integrated,
\begin{equation}
\label{eq:ito12}
\mu(t)=\mu(t_0)+b[W(t)-W(t_0)]-\frac{1}{2}b^2(t-t_0).
\end{equation}
Changing the variable $\mu$ back to $\dot{M}$ gives the analytical solution of
the SDE (\ref{eq:ito10}),
\begin{equation}
\label{eq:ito13}
\dot{M}(t)=\dot{M}(t_0)e^{b[W(t)-W(t_0)]-b^2(t-t_0)/2}.
\end{equation}
Using the fact that $W(t)$ is a normally distributed random variable
with zero mean and variance $\langle W(t)^2\rangle=t$, it is easy to
calculate the first two moments for $\dot{M}$,
\begin{equation}
\label{eq:ito14}
\begin{split}
\langle \dot{M}(t)\rangle &= \langle \dot{M}(t_0)\rangle,\\
\langle \dot{M}(t)^2\rangle &= \langle \dot{M}(t_0)^2\rangle e^{b^2(t-t_0)} .
\end{split}
\end{equation}
In other words, if $\dd\dot{M}/\dot{M}$ is a white-noise process, then
the mean of $\dot{M}$ remains constant, whereas the variance grows
exponentially.

Note that according to equation (\ref{eq:ito13}) the mass deposition
rate is non-negative, $\dot{M}\geq0$.  The limiting value $\dot{M}=0$
corresponds to the stable situation, when the heating rate equals the
cooling rate exactly.  According to the equation (\ref{eq:toy3}) if
$\dot{M}=0$ the heating rate is given by
$H=L_\text{inner}+L_\text{outer}$.  In the present model, however, the
heating ability of AGN is explicitly limited to the inner part of the
cluster.  The maximum heating rate is, therefore, $H=L_\text{inner}$.
If $L_\text{outer}\neq0$ the mass deposition rate is always positive,
$\dot{M}>0$, and the cold gas continuously accumulates in the
cluster's centre.

In order to understand how the stochastic behaviour of $\dot{M}$
influences the energy balance in the ICM, it is necessary to make a
further assumption about the reaction of the ICM to the change in
$\dot{M}$. As was shown above, see (\ref{eq:toy7}), in the state of
the stable deterministic equilibrium the enthalpy of the ICM,
$\omega$, is inversely proportional to $\dot{M}$. By presuming that
this is also true in the case of the stochastically behaving
$\dot{M}$, the reaction of the enthalpy to the change in $\dot{M}$ can
be calculated using the rules of It\^{o} calculus,
\begin{equation}
\label{eq:ito15}
\begin{split}
\dd\omega &=
-L_\text{outer}\frac{1}{\dot{M}^2}\dd\dot{M}
+L_\text{outer}\frac{1}{\dot{M}^3}(\dd\dot{M})^2\\
&=-\frac{L_\text{outer}}{\dot{M}}\left(b\dd W-b^2\dd t\right).
\end{split}
\end{equation}
Substitution of equations (\ref{eq:ito15}) and (\ref{eq:ito10}) into
equation (\ref{eq:toy5}) yields the following SDE for the ICM heating,
\begin{equation}
\label{eq:ito16}
\dd H(t) = [H(t)-L_\text{inner}(t)]b\dd W(t) - b^2L_\text{outer}\dd t.
\end{equation}
%
\section{Heating process}
\label{sec:heat}

Considering for simplicity that $L_\text{inner}\approx$~const, the heating SDE
(\ref{eq:ito16}) can be rewritten with the dimensionless variables:
$h=H/L_\text{inner}$, $\lambda=L_\text{outer}/L_\text{inner}$, and $\tau=b^2t$,
\begin{equation}
\label{eq:heat17}
\dd h(\tau) = [h(\tau)-1]\dd W(\tau) - \lambda\dd\tau.
\end{equation}

\begin{figure}
\centering\includegraphics[width=0.95\linewidth]{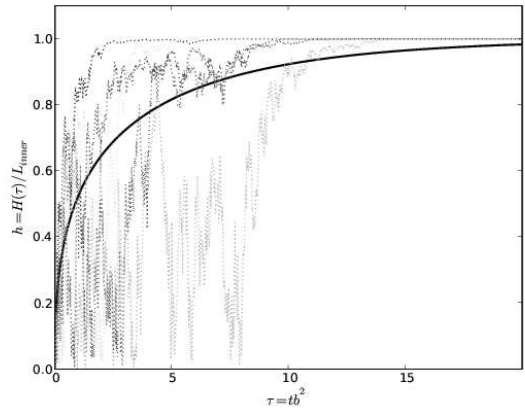}
\caption{Five sample heating curves (dotted lines) produced using equation
(\ref{eq:heat21}) with $h_0=0.1$, and the theoretical mean of the heating process
(black line) as given by equation (\ref{eq:heat21}).  Note the change of
the variability (dispersion) with time, compared to Fig.~\ref{fig:heat2a}.}
\label{fig:heat2}
\end{figure}

To solve the equation (\ref{eq:heat17}) it is best to start by letting
$\lambda=0$. This corresponds to the scenario when the AGN can heat the entire
cluster.  The change of the variable $x=\ln(1-h)$ leads to the following
equation,
\begin{equation}
\label{eq:heat18}
\begin{split}
\dd x &= -\frac{1}{1-h}\dd h - \frac{1}{2(1-h)^2}(\dd h)^2\\
&= \dd W(\tau)-\frac{1}{2}\dd\tau,
\end{split}
\end{equation}
which can be easily integrated, giving the solution,
\begin{equation}
\label{eq:heat19}
h(\tau) = 1 - (1-h_0)e^{W(\tau)-\tau/2},
\end{equation}
where $\tau_0=0$, $h_0=h(0)$, $W(0)=0$.  It is important to note, however, that
although the change of variable guarantees that $h<1$ it also allows an
unphysical situation when $h<0$. To exclude solutions with the negative heating
it is necessary to treat $h=0$ as a boundary condition.  Clusters that reach this
boundary can be reflected at it ($h$ turns positive), they can be absorbed at the
boundary ($h$ remains zero), or, possibly, a fraction of clusters can be
reflected and the remaining part gets absorbed.  The reflective boundary seems to
be a natural choice, since the case of the absorbing boundary requires permanent
extinction of the {\it active} nucleus, while generally $\dot{M}\neq0$, leading
to continuous accumulation of the cold gas in the centre. According to the
equation (\ref{eq:heat19}) the heating remains positive in the range
$W(\tau)<W'=\tau/2-\ln(1-h_0)$. Solving (\ref{eq:heat19}) for $W$, and reflecting
the function around the point $W'$, gives the equation for the reflected process,
valid in the range $W>W'$.  The full solution of the heating SDE, with reflection
at $h=0$ ($W=W'$), is given by,
\begin{equation}
\label{eq:heat20}
h(\tau) = 
\begin{cases}
1 - (1-h_0)e^{W(\tau)-\tau/2}, & W(\tau)<W',\\
1 - \frac{1}{1-h_0}e^{-W(\tau)+\tau/2}, & W(\tau)>W'.
\end{cases}
\end{equation}
Using the solution (\ref{eq:heat20}) it is straightforward to
calculate the mean value of the heating,
\begin{equation}
\label{eq:heat21}
\begin{split}
\langle h(\tau)\rangle &= 1 -
\frac{h_0}{2}
\text{erfc}\left[\frac{\tau/2+\ln(1-h_0)}{\sqrt{2\tau}}\right]\\
&-\frac{h_0e^\tau}{2(1-h_0)}
\text{erfc}\left[\frac{3\tau/2+\ln(1-h_0)}{\sqrt{2\tau}}\right],
\end{split}
\end{equation}
which shows that the equilibrium $h\rightarrow1$ ($H\rightarrow L$) is
approached exponentially quickly as $\tau\rightarrow\infty$, see
Fig.\ref{fig:heat2}.  Using (\ref{eq:heat20}) it is also possible to
obtain analytical formulae for the variance and the autocorrelation
function, but the expressions are excessively long and will not be
reproduced here.

\begin{figure}
\centering\includegraphics[width=0.95\linewidth]{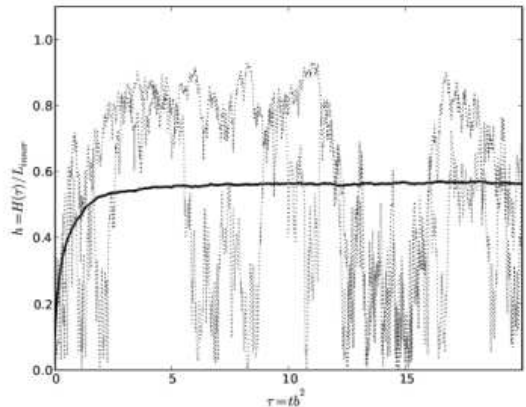}
\caption{Two sample heating curves (dotted lines) produced using
  numerical solutions of equation (\ref{eq:heat17}) with $h_0=0.1$,
  $\lambda=0.2$, and the mean of the heating process (black line)
  computed as an average of 10000 sample heating curves.  Note that
  the amount of the variability (dispersion) remains constant in time,
  unlike in the case that $\lambda=0$, see Fig.~\ref{fig:heat2}.}
\label{fig:heat2a}
\end{figure}

The analysis of the solutions for the SDE can be complemented by
considering the probability density function (PDF) of the
processes. This is of interest for building a framework with which to
interpret observations. For example, the most probable value of the
observed heating rate is the modal value, where the PDF
peaks. However, if the PDF is skewed, the mode might provide a poor
estimate for the long term average heating rate, which is given by the
mean of the distribution.

\begin{figure}
\centering\includegraphics[width=0.95\linewidth]{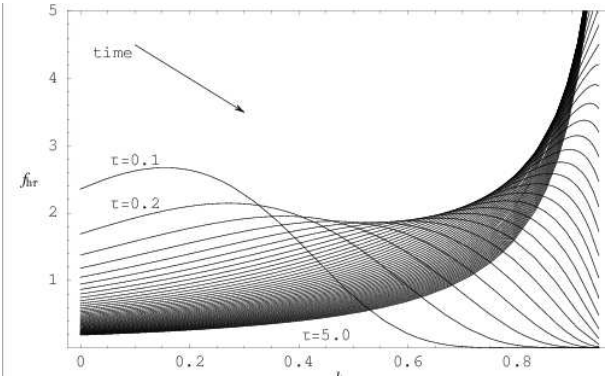}
\caption{Evolution of the PDF for the heating process with $\lambda=0$, which
started at $\tau=0$ as a delta function, $\delta(h-0.1)$. The curves
correspond to the times $\tau=0.1,\ldots,5$ with the step $\Delta\tau=0.1$.}
\label{fig:heat3}
\end{figure}

The stochastic process described by the SDE (\ref{eq:heat17}) has a
PDF $f_{h\tau}(h,\tau)$, which satisfies the following Fokker-Plank
equation (FPE),
\begin{equation}
\label{eq:heat22}
\pder{f_{h\tau}}{\tau} = \lambda\pder{f_{h\tau}}{h}
+\frac{1}{2}\pder{^2}{h^2}\left[(h-1)^2f_{h\tau}\right],
\end{equation}
\citep[see][for a general derivation of the relation between FPE and
SDE]{Gardiner04}.  Substitution of the variable $x=\ln(1-h)$ and the
corresponding change the PDF function $f_{h\tau}=f_{x\tau}e^{-x}$ (which
preserves the probability measure) transforms the FPE (\ref{eq:heat22}) into the
equation
\begin{equation}
\label{eq:heat23}
\pder{f_{x\tau}}{\tau} = 
\pder{}{x}\left[ \left(\frac{1}{2}-\lambda e^{-x}\right) f_{x\tau}\right]
+\frac{1}{2}\pder{^2f_{x\tau}}{x^2},
\end{equation}
which describes a diffusion process on the interval\footnote{Square
  bracket means that the end of the interval is closed, and the round
  bracket that that it is open.} $x\in(-\infty,0]$ in a medium with
  constant diffusion coefficient, $1/2$, in presence of the
  inhomogeneous force field, $1/2-\lambda e^{-x}$.  The reflective
  boundary condition at $x=0$ can be satisfied automatically by
  extending the solution interval to the whole real axis, and
  reflecting the distribution at $0$ so that it becomes an even
  function.  In the case where $\lambda=0$, the equation can be easily
  solved,
\begin{equation}
\label{eq:heat24}
f_{x\tau} = \frac{e^{x/2-\tau/8}}{\sqrt{2\pi\tau}}
\int_{-\infty}^{\infty}
\left(e^{-(x-\xi)^2/2\tau}+e^{-(x+\xi)^2/2\tau}\right)
f_{\xi0}e^{\xi/2}\dd\xi,
\end{equation}
where $f_{\xi0}$ is the initial ($\tau=0$) PDF.  An example of how the
PDF changes with time, starting from $f_{\xi0}\propto\delta(\xi-0.1)$,
is shown in Fig.~\ref{fig:heat3}.  For large values of $\tau$, the PDF
becomes a narrow peak at $h=1$, in agreement with the SDE result,
$\langle h\rangle\rightarrow1$, see (\ref{eq:heat21}) and
Fig.~\ref{fig:heat2}.

The connection between the FPE and the SDE also helps to explain the
role of the $(h-1)\dd W$ term in the original SDE (\ref{eq:heat17}). If the
term was simply $\dd W$ the resulting FPE would be an equation of homogeneous
diffusion. This would correspond to a case in which the feedback is
not scaled with $\dot{M}$. The limiting distribution in this case
would be a uniform one. The $h-1$ factor creates a force term in the
equation, ensuring that, at all times, diffusion in the direction of
$h=1$ is preferred over the opposite direction, no matter what kind of
the initial distribution is assumed.

The FPE (\ref{eq:heat22}) can be rewritten in the form of a
conservation law, $\partial_\tau f_{h\tau}=\partial_hJ_{h\tau}$, the
function $J_{h\tau}$ is called the probability flux function. In the
stationary regime, the time derivative is zero, $\partial_\tau
f_{h\tau}=0$.  It describes a statistically stable scenario, which is
presumably reached in the limit of large $\tau$.  It follows that in
this case $J_{h\tau}=\text{const}$.  Because of the reflective
boundary at $h=0$ the fluxes on the left and right of this point must
have opposite signs $J_{0t}=-J_{0t}=0$, and therefore in the
stationary case the probability flux is zero.  $J_{h\tau}=0$ is an
ordinary differential equation, which can be readily solved,
\begin{equation}
\label{eq:heat25}
f_h \propto
\begin{cases}
\frac{1}{(1-h)^2}, &\lambda=0,\\
\frac{1}{(1-h)^2}e^{-2\lambda/(1-h)}, &\lambda\neq0.
\end{cases}
\end{equation}
\begin{figure}
\centering\includegraphics[width=0.9\linewidth]{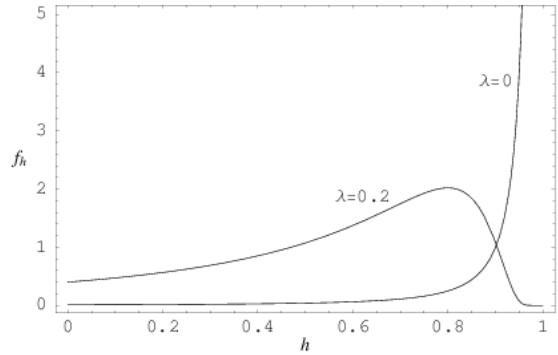}
\caption{Asymptotic ($\tau\rightarrow\infty$) shapes of the PDFs of
  the cluster heating according to the equations (\ref{eq:heat25}).  If
  $\lambda\neq0$ the PDF has a maximum at $1-\lambda$ (in this case
  $\lambda=0.2$). For comparison both PDFs were normalised on the
  interval $[0,0.95]$.}
\label{fig:heat4}
\end{figure}
In the case $\lambda\neq0$ the normalised distribution is given by,
\begin{equation}
\label{eq:heat25a}
f_h = \frac{2\lambda}{(1-h)^2}e^{-2\lambda\frac{h}{1-h}}.
\end{equation}
In the case $\lambda=0$ the function $f_h$ is not normalisable on $[0,1]$.
However, the solutions (\ref{eq:heat25}) are still useful for understanding the
qualitative distinction between the two cases. If $\lambda\neq0$ the PDF has a
peak at $h=1-\lambda$, see Fig.~\ref{fig:heat4}, suggesting that clusters in this
case would appear to be under-heated, on average. The change in the character of
the PDF is due to the inhomogeneous force $1/2-\lambda/(1-h)$ in the diffusion
equation (\ref{eq:heat23}).  Along with the constant term $1/2$, which forces
diffusion in the direction of $h=1$, there is a counteracting term,
$-\lambda/(1-h)$, which reverses the diffusion on the interval $1-2\lambda<h<1$.

\begin{figure}
\centering\includegraphics[width=\linewidth]{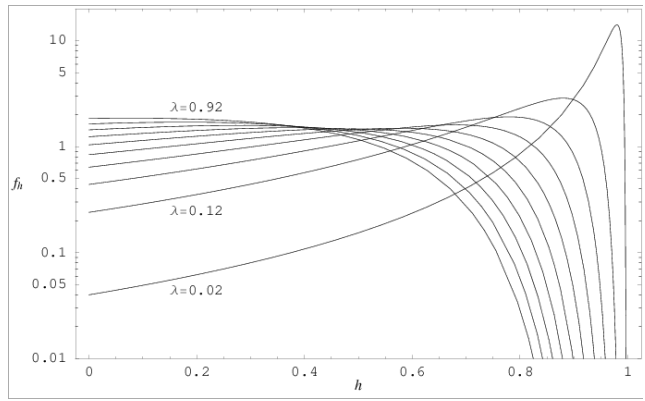}
\caption{Normalised PDFs in the case $\lambda\neq0$ as given by the equation
(\ref{eq:heat25a}).  The curves correspond to the values of
$\lambda=0.02,\ldots,0.92$ with the step $0.1$.}
\label{fig:heat5}
\end{figure}

A physical interpretation of this property becomes transparent in the context of
the toy model.  The ability of the AGN to heat the cluster is limited to the 
volume $V_\text{inner}$ (with X-ray luminosity $L_\text{inner}$ or 1).  The
additional mass deposition from the volume $V_\text{outer}$ (with the
luminosity $L_\text{outer}$ or $\lambda$) can not be counterbalanced by the AGN heating.
This results in a shift of the balance from the exact match, $h=1+\lambda$, to
the lower end with the most likely value $h=1-\lambda$
($H=L_\text{inner}-L_\text{outer}$).  The shape of the PDF is asymmetric and
varies with $\lambda$, see Fig.~\ref{fig:heat5}.  Also the mode or the most
probable heating rate is different from the mean or the average heating rate in
this case, as the distribution is skewed.

It is possible to find an analytical solution to the SDE (\ref{eq:heat17})
in the case $\lambda\neq0$,
\begin{equation}
\label{eq:heat26}
h(\tau) = 1 - \varepsilon(\tau) \left( 1 - h_0 + \lambda
\int_0^\tau\frac{\dd\tau'}{\varepsilon(\tau')} \right),
\end{equation}
where $\varepsilon(\tau)=e^{W(\tau)-\tau/2}$ is known as the exponential
martingale.  This solution must be completed with the reflected process, so that
$h(\tau)>0$.  However, because the analytical quadrature of the integral term is
unknown \citep[see, e.g.,][for details]{Goodman06}, this can not be done as
easily as in the case $\lambda=0$.

The long-term behaviour of the process in the case $\lambda\neq0$ was
already studied using the approach based on the FPE (\ref{eq:heat23}).
The sample heating curves can be calculated using numerical solutions
of the SDE (\ref{eq:heat17}). In Fig.~\ref{fig:heat2a} two sample
curves are plotted for processes with $\lambda=0.2$, starting from
$h_0=0.1$.  The mean of the heating rapidly approaches the value
$\avrg{h}\approx0.55$ (note that $h=0.8$ is the most probable value)
while contrary to the $\lambda=0$ case the standard deviation remains
finite, and the sample curves exhibit a lot of variation.

Note that the balance of heating and cooling in the case when
$\lambda\neq0$ is given by $h=1+\lambda$ (not $h=1$) and therefore it
is never reached.  This is a direct consequence of the imposed limit
on the AGN ``fuel supply'' (i.e., its heating capability, see
section~\ref{sec:toy}) to the region $V_\text{inner}$.  The mass
deposited from the $V_\text{outer}$ is not a part of the AGN feedback
cycle, and accumulates in the centre.  In practice this sets a limit
to the value of $\lambda$, which could be determined from future
cluster surveys.
 
\section{Discussion}
\label{sec:discuss}
In the context of the model developed here, it is not surprising that most
observed clusters are found to have a very tight correspondence between the
heating power and the radiative cooling rate \citep{Dunn08}.  While it would be
naive to equate the heating rate in the current model with the power input
measured from the size of the cavities in the clusters, it is reasonable to think
that the two energy rates are positively correlated.

In the presented model the self-tuning of the energy balance is rapid,
exponentially quick, in fact $\propto e^{\tau}$ (see equation
(\ref{eq:heat21})).  It works even in the case when the AGN is able to
heat only a fraction of the cluster volume (see
Fig.~\ref{fig:heat2a}).  This shows that the correspondence of the
cooling power and the AGN heating seen in the observations may indeed
be a general rule, which is characteristic of the feedback process.
Since the underlying PDF of the heating can be skewed (see
Fig.~\ref{fig:heat5} and equation (\ref{eq:heat25a})) it is worth
noting that the heating rates inferred from the observations most
frequently must lie near the mode of the distribution, and can not be
used directly to find the mean or the average heating rate. The true
picture can only be uncovered in a complete survey.


As was shown above, the time dependent PDF becomes very close to the
static solution for times $\tau>10$, see Figs. \ref{fig:heat2} and
\ref{fig:heat2a}. The physical time is proportional to $\tau$ with the
proportionality constant $b^{-2}$ (see the normalisation of variables
just before the equation (\ref{eq:heat17})). The physical time limit
is, therefore, $t>10b^{-2}$, which puts the lower limit on the model
parameter $b$. If the physical time is of the order of the Hubble time
the limit on $b$ is given by $b>3\times10^{-5}$ yr$^{-1/2}$.  In 
principle, this value can be verified observationally using the
relation between the variance of $\dot{M}$ at different times, as
given by the equation (\ref{eq:ito14}). Given sufficient statistics
for the mass deposition rates across the population of clusters in
different redshift bins, the prediction of the exponential growth of
the $\langle\dot{M}(t)^2\rangle$ can be tested.  The values of the model
parameters $b$ and $\lambda$ can also be found.  It is unlikely that,
over the entire lifetime of a cluster, the real distribution of the
heating is very close to the $\lambda=0$ case. Since in this case the
AGN would, on average, prevent any cooling flow from developing and
also prevent any star formation, which, in reality, is observed to be
enhanced in some brightest cluster galaxies \citep{Rafferty06}.  This
leaves the single parameter distribution (\ref{eq:heat25a}) as a plausible
heating PDF for the population of ``mature" (of age $\tau>10$) clusters.

The current model sketches just one possible scenario for the
feedback.  In order to make the model robust and simple, many
assumptions were made: the insignificance of the magnetic field and
the kinetic energy, rapid thermalisation of the AGN energy, and the
inverse proportionality between the enthalpy and the mass flow.  They
have to be tested and verified in the future using numerical
simulations and analysis of the observations. In addition, it is quite
possible, for example, that there can be a delay in the AGN response
to the cooling flow \citep{Pope07}.  During this delay an excess of
cold material could accumulate around the AGN.  This would be
equivalent to effectively growing the cluster in size from the point
of view of the feedback loop. AGN heating of such a larger cluster
could then result in the overheating of the ICM, and possibly act as a
destabilising factor.

Despite the limitations of the model, this work demonstrates that the
employment of stochastic calculus appears to be particularly well
suited for the analysis of AGN feedback.  The combination of It\^o
calculus and the Fokker-Plank equations provides a simple, yet
powerful, way of investigating behaviour of the time-dependent
variability of AGN feedback in clusters.

\section{Acknowledgements}
GP would like to thank Ian McHardy and Tom Maccarone for useful
discussions of parts of this work and STFC for financial support
though a rolling grant.  ECDP thanks the Department of Foreign Affairs
and International Trade for funding through a Government of Canada
Post-Doctoral Research Fellowship, CITA for funding through a National
Fellowship and also Arif Babul for additional funding.  This research
has made use of NASA's Astrophysics Data System Bibliographic
Services.

\bibliography{gbp}

\end{document}